\documentclass[twocolumn,showpacs,amsmath,amstex,amssymb,mathfonts,prb]{revtex4-1}

\usepackage{graphicx,bm,units}

\begin{document}

\title{Manifestly gauge independent formulations of the ${\bm Z}_2$ invariants}

\author{Emil Prodan}
\address{Department of Physics, Yeshiva University, New York, NY 10016} 

\begin{abstract} We use a``monodromy" argument to derive new expressions for the ${\bm Z}_2$ invariants of topological insulators with time-reversal symmetry in 2 and 3 dimensions. The derivations and the final expressions do not require any gauge choice and the calculation of the invariants is based entirely on the projectors onto the occupied states. Explicit numerical tests for tight-binding models with strongly broken inversion symmetry are presented in 2 and 3-dimensions.
\end{abstract}

\pacs{73.43.-f, 72.25.Hg, 73.61.Wp, 85.75.-d}

\date{\today}

\maketitle

\section{Introduction}

Topological insulators represent a new class of materials where the topology of the bulk electronic structure induces non-trivial effects such as the emergence of edge states.\cite{HALDANE:1988rh,Kane:2005np,Kane:2005zw} The edge states are robust against smooth deformations of the crystals or the presence of disorder.\cite{Prodan:2009lo,Prodan:2009mi}

Among all known classes of topological insulators, the time-reversal invariant ones have a special status because they were already engineered and characterized in laboratories.\cite{Bernevig:2006hl,Koenig:2007ko,Fu:2007vs,Hsieh:2008vm} As pointed out early in the development of the field, the bulk-edge correspondence in time-reversal invariant insulators obeys a ${\bm Z}_2$ classification.\cite{Kane:2005np} It was also clear from early stages that the topology of the time-reversal invariant bulk electronic structures is classified  by the twisted Real K-Theory,\cite{Kane:2005zw} which pointed again to a ${\bm Z}_2$ classification. Homotopy arguments lead to the same conclusion.\cite{Moore:2007ew}

Several explicit formulations of the ${\bm Z}_2$ invariants were given along the years but all of them involve globally smooth gauges. Gauge independent invariants were formulated in Ref.~\onlinecite{Loring2010vy}, but their effectiveness remains to be tested in 3 dimensions. For time-reversal invariant systems, which inherently have trivial Chern numbers, an important result by Panati assures the existence of such globally smooth gauges.\cite{Panati:2007sy} Unfortunately, the result by Panati is not constructive and at this point we don't have a standard algorithm to construct globally smooth gauges, something that in many instances proved to be a formidable task. 

The early work of Ref.~\onlinecite{Kane:2005zw} proposed to look at the Pfaffian of a particular overlap matrix as function of the $k$-vector. Generically, this Pfaffian can become zero at isolated $k$-points, which always come in pair. The number of paired first order zeroes, taken modulo 2, was found to be a topological invariant. Several equivalent expressions of the ${\bm Z}_2$ invariant were derived in Ref.~\onlinecite{Fu:2006ka}. This work introduced the notion of time-reversal polarization, which was shown to be quantized modulo 2. The computation of the time-reversal polarization requires a globally smooth gauge, which must also be adapted to the time-reversal symmetry (see Eq.~3.10). The time-reversal polarization approach inspired yet another formulation of the invariant, involving the Pfaffian and the square root of the determinant of another overlap matrix, computed at the time-reversal invariant $k$-points. This formulation played a special role since it admitted extentions to higher dimensions.\cite{Fu:2007ti,Fu:2007vs} Furthermore, the ${\bm Z}_2$ invariant was formulated as an obstruction against achieving a globally smooth gauge of certain kind, leading to yet another equivalent expression involving the Berry curvature and the Berry phase. Starting from this later expression of the invariant, Fukui et al were able to develop a (almost) gauge-independent method of calculus in 2-dimensions (the method still requires a time-reversal adapted gauge at the boundary of half of the Brillouin zone).\cite{FukuiJPSJ2005gu} This method was later extended in 3-dimensions,\cite{FukuiJPSJ2007ny} and it became the method of choice when computing the ${\bm Z}_2$ invariants for non centro-symmetric systems.\cite{Essin:2007ij,XiaPRL2010ni,FengPRB2010ry,SoluyanovPRB2011bt,FengPRL2010gu,WadaPRB2011fu}   

The requirement of special smooth gauges in the classic formulations of the ${\bm Z}_2$ invariants is unfortunate, and the (almost) gauge-independent method of calculus developed by Fukui et al can be quite involved. For example, the prediction of the first topological insulators in 3D was in great part possible because the ${\bm Z}_2$ invariants simplify tremendously when inversion symmetry is also present.\cite{Fu:2007vs} Without inversion symmetry, the evaluation of the ${\bm Z}_2$ invariants remains a very difficult task. For example, in a study on strained bulk HgTe material,\cite{DaiPRB2008ji} a non centro-symmetric system, even with a tight-binding model it was more convenient to complete slab calculations and look directly at the surface states rather than compute the bulk ${\bm Z}_2$ strong invariant. A similar approach was followed in a recent first-principle study on the non centro-symmetric metacinnabar compound.\cite{Virot2010by} So evaluating the ${\bm Z}_2$ invariant is already difficult at the level of tight-binding modeling, but the difficulty becomes overwhelming when attempting first principle electronic structure calculations. This aspect was recently discussed in Ref.~\onlinecite{Soluyanov2011gy}, where a solution was proposed based on hybrid Wannier functions. Subsequent work,\cite{Yu2011re} has also employed hybrid Wannier functions to derive equivalent ${\bm Z}_2$ invariants in 2 dimensions. Notably, this later work made use, like us, of the full (not just the trace) adiabatic connection. The use of hybrid Wannier functions to efficiently re-formulate the topological invariants was originally introduced in Ref.~\onlinecite{Ringel2010vo}.  

In 3 dimensions, topological insulators were also shown to display quantized magneto-electric polarization,\cite{QiPRB2008ng} which can be written as a Chern-Simons integral. This invariant was shown to be completely equivalent to the previously introduced strong ${\bm Z}_2$ invariant.\cite{Wang2010xs} The computation of the Chern-Simons integral requires again a globally smooth gauge and the fundamental difficulties introduced by this requirement were already highlighted in Ref.~\onlinecite{CohPRB2011rt}. To date, nobody has achieved a direct evaluation of this Chern-Simons integral, even for tight-binding models. The magneto-electric polarization was computed indirectly, using the second Chern number and dimension reduction technique.\cite{Qi:2008cg,Hughes2010gh} It became a sure fact that computing the second Chern number in 4 dimension is much easier than computing the Chern-Simons integral in 3 dimensions, and this is precisely because the second Chern number admits a manifestly gauge indepent expression based entirely on the projector onto the occupied states.   

Our present work provides equivalent formulations of the ${\bm Z}_2$ bulk invariants that are manifestly gauge independent. We report results for both 2 and 3 dimensions. The new formulas are computationally trivial for both tight-binding and first principle approaches. The key to these results is a ``monodromy" argument, which was recently introduced in Ref.~\onlinecite{Hughes2010gh}. Basically, instead of looking at the polarization, we examine the full non-abelian adiabatic transport along time-reversal invariant lines in the Brillouin zone and take advantage of the special behavior under the time-reversal operation. Using the elementary properties of the determinants and Pfaffians, we are able to show that the determinant of the monodromy, computed along closed time-reversal invariant paths in the Brillouin torus, can be written as the square of a well defined quantity. This quantity divided by the square root of the determinant of the monodromy takes the quantized values of $\pm 1$, and becomes the building block for our invariants. 

In 2 and 3-dimensions, we look at pairs of time-reversal paths on the Brillouin torus. For such pairs, we show that the square root of the determinants of the monodromies can be taken in a canonical way, allowing us to define a true ${\bm Z}_2$ topological invariant for each such pair. In 2-dimensions, this construction gives the unique ${\bm Z}_2$ invariant, while in 3-dimensions it generates four independent weak invariants, which can be subsequently used to generate the strong ${\bm Z}_2$ invariant. 

We use tight-binding models with time-reversal symmetry to test our formulations and to show how the construction works. The models include interactions that strongly break the inversion symmetry.

\section{The main construction}

Let us consider a one dimensional, translational invariant lattice model, described by a Bloch Hamiltonian $H_k$, a $N \times N$ complex matrix with $k$-dependent entries. The Hamiltonian acts on the fixed space of $N$-component complex vectors, the ${\bm C}^N$ space. We assume time-reversal symmetry, that is, we assume the existence of an antilinear operator $\theta$ acting on ${\bm C}^N$, such that:
\begin{equation}
\theta H_k \theta^{-1} = H_{-k}.
\end{equation}
We also assume that we are dealing with an insulator, so $H_k$ is assumed to have a spectral gap at all $k$'s. We denote by $P_k$ the projector onto the states below this spectral gap.

Our construction starts from the monodromy $U_{k,k'}$ defined as the unique solution to the following differential equation:
\begin{equation}\label{AdTransp}
\begin{array}{c}
i\frac{d}{d k} U_{k,k'} =i [P_k,\partial_k P_k] U_{k,k'},
\end{array}
\end{equation}
with the initial condition $U_{k',k'}$=$P_{k'}$.  Here, $k'$ is an arbitrary but fixed $k$-point. The monodromy provides a parallel transport, that is, an isometric mapping of the space $P_{k'}{\bm C}^N$ into the space $P_k{\bm C}^K$, more precisely:
\begin{equation}
P_k = U_{k,k'} P_{k'} U_{k,k'}^{-1}.
\end{equation}
The monodromy is also known to generate a one parameter unitary group:
\begin{equation}
U_{k,k'}U_{k',k''}=U_{k,k''}, \ \ U_{k,k'}U_{k',k}=Id.
\end{equation}

Eq.~\ref{AdTransp} can be derived, and it was first derived (see Ref.~\onlinecite{simon1983}), from the modern formulation of the Adiabatic Theorem.\cite{Nenciu:1981kx} If a local gauge (i.e. smoothly k-dependent bases for $P_k{\bm C}^N$ spaces) was pre-defined, then Eq.~\ref{AdTransp} takes the more familiar form:
\begin{equation}\label{Will}
\begin{array}{c}
\frac{d}{dk}\hat{U}(k)=i\hat{A}(k) \hat{U}(k),
\end{array}
\end{equation}
where $\hat{A}(k)$ is the full non-abelian adiabatic connection discussed by Wilczek and Zee in Ref.~\onlinecite{wilczek:1984bs}. For a more detailed discussion one can consult Ref.~\onlinecite{Prodan:2009hg}. We will, however, want to stay way from the later Eq.~\ref{Will} because a smooth gauge can be, in general, chosen only locally. And even though time-reversal invariant systems admit global smooth gauges,\cite{Panati:2007sy} constructing such a globally smooth gauge can be quite a formidable task.

One key observation is that Eq.~\ref{AdTransp} can be integrated without making use of any gauge. Indeed, assume that we divided the interval $[k',k]$ in small subintervals: $k'=k_1 \ldots <k_n = k$, and let us form the product:
\begin{equation}\label{Solution}
U_i=P_{k_i} P_{k_{i-1}} \ldots P_{k_1}.
\end{equation}
By a simple term counting, one can easily see that $U_i$ satisfies the equation:
\begin{equation}
\begin{array}{c}
i(U_i - U_{i-1})\medskip \\
=i\{(P_{k_i}-P_{k_{i-1}})P_{k_{i-1}}-P_{k_i}(P_{k_i}-P_{k_{i-1}})\}U_{i-1}
\end{array}
\end{equation}
But this equation is nothing else but the finite difference version of our original Eq.~\ref{AdTransp}. In other words, Eq.~\ref{AdTransp} can be integrated by forming the sequenced product shown in Eq.~\ref{Solution}, using a fine-enough finite difference step. This discussion is not limited to one dimension but it can be applied to the parallel transport along any arbitrary path in higher dimensional Brillouin zones. On a more technical note,  let us state that the projectors $P_k$ can be computed without using any particular gauge. There is quite a substantial number of different ways to accomplish that,  but just for the sake of explicitness, let us mention that the projector onto a particular eigenvalue $\epsilon_i(k)$ can be computed as:
\begin{equation}
P_{\epsilon_i(k)}=F_i(H_k),
\end{equation}
where $F_i$ is the so called interpolating polynomial defined by $F_i(\epsilon_j)=\delta_{ij}$.

Now let us conjugate Eq.~\ref{AdTransp} by $\theta$:
\begin{equation}
\begin{array}{c}
\theta\{i\frac{d}{d k} U_{k,k'}\}\theta^{-1} =\theta\{ i [P_k,\partial_k P_k] U_{k,k'}\} \theta^{-1},
\end{array}
\end{equation}
which leads to:
\begin{equation}
\begin{array}{c}
i\frac{d}{d k} (\theta U_{k,k'} \theta^{-1}) =i  [P_{-k},\partial_k P_{-k}] (\theta U_{k,k'} \theta^{-1}).
\end{array}
\end{equation}
Also, $\theta U_{k',k'} \theta^{-1}$ becomes the identity on $P_{-k'}{\bm C}^N$. This leads us to conclude that:
\begin{equation}
\theta U_{k,k'} \theta^{-1}=U_{-k,-k'}.
\end{equation}

Therefore, if we want to compute the monodromy from $-\pi$ to $\pi$, we can write:
\begin{equation}
\begin{array}{c}
U_{\pi,-\pi}=U_{\pi,0}U_{0,-\pi} \medskip \\
=U_{\pi,0}\theta U_{0,\pi} \theta^{-1}=U_{\pi,0}\theta U_{\pi,0}^{-1} \theta^{-1}.
\end{array}
\end{equation}
The monodromy $U_{\pi,-\pi}$ maps the space $P_{\pi}{\bm C}^N$ into itself. We can therefore enquire about the determinant of this monodromy. It is a fact that the determinant of $U_{\pi,-\pi}$ is always equal to one for systems with time-reversal and inversion symmetries. If the inversion symmetry is broken, this determinant can take, in principle, any value on the unit circle. As we shall see in the following, for the most general situation, the determinant can be written as the square of a well defined quantity, which will become the building block for our ${\bm Z}_2$ invariants.

To see this, let us choose an arbitrary basis in $P_\pi {\bm C}^N$ and $P_0 {\bm C}^N$, which we denote by $\{e^\pi_\alpha\}$ and $\{e^0_\alpha\}$, respectively. Before we start the calculation, let us point the following fact ($Q$ = arbitrary linear operator acting on $P_0{\bm C}^N$):
\begin{equation}
\begin{array}{c}
\langle e^0_\alpha|\theta Q | e^0_\beta \rangle = \langle e^0_\alpha|\theta (\sum_\delta \langle e^0_\delta |Q|e^0_\beta \rangle |e^0_\delta\rangle ) \medskip \\
=\sum_\delta \overline{\langle e^0_\delta |Q|e^0_\beta \rangle} \langle e^0_\alpha|\theta |e^0_\delta \rangle.
\end{array}
\end{equation} 
We can put the above fact in a more convenient form,
\begin{equation}\label{AntiLin}
\begin{array}{c}
\langle e^0_\alpha|\theta Q | e^0_\beta \rangle =\sum_\delta \langle e^0_\alpha|\theta |e^0_\delta \rangle \overline{\langle e^0_\delta |Q|e^0_\beta \rangle},
\end{array}
\end{equation}
in which case we see a simple rule, that when inserting an identity operator $\sum_\delta |e^0_\delta\rangle \langle e^0_\delta|$ after the anti-linear operator $\theta$, all the resulting matrix elements after $\theta$ must be complex conjugated.

We can now start the calculation:
\begin{equation}
\begin{array}{c}
\langle e^\pi_\alpha|U_{\pi,-\pi}|e^\pi_\beta\rangle = \langle e^\pi_\alpha|U_{\pi,0}\theta U_{\pi,0}^{-1} \theta^{-1}|e^\pi_\beta\rangle \medskip \\
=\langle e^\pi_\alpha|U_{\pi,0}|e^0_\delta\rangle \langle e^0_\delta|\theta |e^0_\gamma \rangle \overline{\langle e^0_\gamma| U_{\pi,0}^{-1}|e^\pi_\xi \rangle } \overline {\langle e^\pi_\xi | \theta^{-1}|e^\pi_\beta\rangle }.
\end{array}
\end{equation}
Summation over repeating indices was assumed above. We denote by $\hat{U}$ the matrix of elements
\begin{equation}
\hat{U}_{\alpha \beta}=\langle e^\pi_\alpha|U_{\pi,0}|e^0_\delta\rangle.
\end{equation}
Note that
\begin{equation}
\langle e^0_\gamma| U_{\pi,0}^{-1}|e^\pi_\xi \rangle = \hat{U}^{-1}_{\gamma \xi}.
\end{equation}
Also, since $U_{\pi,0}$ is an isometry, the matrix $\hat{U}$ is unitary and consequently:
\begin{equation}
\overline{\langle e^0_\gamma| U_{\pi,0}^{-1}|e^\pi_\xi \rangle}=\hat{U}^T_{\gamma \xi}.
\end{equation}

We denote by $\hat{\theta}_{0}$ and $\hat{\theta}_\pi$ the matrices of elements:
\begin{equation}
(\hat{\theta}_0)_{\alpha \beta}=\langle e^0_\alpha|\theta |e^0_\beta \rangle, \ \ (\hat{\theta}_\pi)_{\alpha \beta}=\langle e^\pi_\alpha|\theta |e^\pi_\beta \rangle 
\end{equation}
and we point out the identity:
\begin{equation}
\overline {\langle e^\pi_\xi | \theta^{-1}|e^\pi_\beta\rangle }=(\hat{\theta}^{-1}_\pi)_{\xi \beta},
\end{equation}
where the complex conjugation is due to the property stated in Eq.~\ref{AntiLin}.
With these technicalities behind us, we can now state that:
\begin{equation}
U_{\pi,-\pi} = \hat{U} \hat{\theta}_0 \hat{U}^T \hat{\theta}_\pi^{-1}.
\end{equation}
Therefore:
\begin{equation}
\det \{U_{\pi,-\pi} \} = \det \{ \hat{U} \hat{\theta}_0 \hat{U}^T \hat{\theta}_\pi^{-1} \}
\end{equation}
and, since the $\hat{\theta}$ matrices are antisymmetric, we can use their Pfaffians and the elementary properties of determinants to conclude:
\begin{equation}\label{interm}
\det \{U_{\pi,-\pi} \} =[\mbox{Pf}\{\hat{\theta}_\pi\}^{-1} \det \{ \hat{U} \} \mbox{Pf}\{ \hat{\theta}_0 \}]^2.
\end{equation}
We arrived at our main conclusion:
\begin{equation}\label{Main}
\boxed{\frac{\mbox{Pf}\{\hat{\theta}_\pi\}^{-1} \det \{ \hat{U} \} \mbox{Pf}\{ \hat{\theta}_0 \} }{ \sqrt{\det \{U_{\pi,-\pi}\}} }= \pm 1.}
\end{equation} 
The left hand side of Eq.~\ref{Main} depends on the branch of the square root we chose, but once this choice is made, the value of the left hand side cannot be changed by smooth deformations of the Hamiltonian that keep the insulating gap opened.

One can verify explicitly that the formula is completely independent of the bases chosen at $k$=0 and $k$=$\pi$. Indeed, if we make a change of bases:
\begin{equation}
e^0_\alpha \rightarrow (\hat{W}_0)_{\alpha \beta} e^0_\beta, \ \ e^\pi_\alpha \rightarrow (\hat{W}_\pi)_{\alpha \beta} e^\pi_\beta
\end{equation}
then 
\begin{equation}
\det\{\hat{U}\}\rightarrow \det\{\hat{W}_\pi\} \det \{\hat{U}\} \det\{\hat{W}_0\}^{-1}
\end{equation}
and
\begin{equation}
\begin{array}{c}
\mbox{Pf}\{\hat{\theta}_\pi\} \rightarrow \det \{\hat{W}_\pi\} \mbox{Pf}\{\hat{\theta}_\pi\}, \medskip \\
 \mbox{Pf}\{\hat{\theta}_0\} \rightarrow \det \{\hat{W}_0\} \mbox{Pf}\{\hat{\theta}_0\},
 \end{array}
\end{equation}
so the invariance follows automatically. In fact, Eq.~\ref{Main} can be evaluated without making reference to any basis set, by just using the abstract (fundamental) definition of the determinant and Pfaffian.\cite{SimonTr2005xu}

Because we don't have a canonical way to choose the branch of the square root at the denominator in Eq.~\ref{Main}, we cannot assign a true topological meaning to this formula. For instance, it will be impossible to compare two separate systems, unless we have an explicit way to deform them into each other without closing the direct energy gap. This shortcoming can be eliminated in 2 and 3 dimensions, where true topological ${\bm Z}_2$ invariants can be defined. This is discussed in the following sections.

We should point out that, by using Eq.~\ref{interm} inside the square root,  Eq.~\ref{Main} can be also written as:
\begin{equation}
\frac{\mbox{Pf}\{\hat{\theta}_0\}}{ \sqrt{\det \{\hat{\theta}_0\}}} \left (\frac{\mbox{Pf}\{ \hat{\theta}_\pi \} }{ \sqrt{\det \{\hat{\theta}_\pi\}} }\right )^{-1}= \pm 1,
\end{equation} 
which shows the direct connection between our formulation and Eq.~3.24 of Ref.~\onlinecite{Fu:2006ka}. The only difference is that we provide a different, but equivalent criterion for choosing the branches of the square roots.

\section{The ${\bm Z}_2$ topological invariant in 2-dimensions}

In 2-dimensions, we can follow a pair of time-reversal invariant paths on the Brillouin torus, and generate a pair of quantized numbers like in Eq.~\ref{Main}, such as:
\begin{equation}\label{Xi1}
\Xi_{0}=\frac{\mbox{Pf}\{\hat{\theta}_{(0,\pi)}\}^{-1} \det \{ \hat{U}_0 \} \mbox{Pf}\{ \hat{\theta}_{(0,0)} \} }{ \sqrt{\det \{U_{(0,\pi),(0,-\pi)}\}}}
\end{equation}
for the path 
\begin{equation}\label{path1}
{\bm k}=\ (0,-\pi) \rightarrow {\bm k}=(0,\pi),
\end{equation}
and
\begin{equation}\label{Xi2}
\Xi_{\pi}=\frac{\mbox{Pf}\{\hat{\theta}_{(\pi,\pi)}\}^{-1} \det \{ \hat{U}_\pi \} \mbox{Pf}\{ \hat{\theta}_{(\pi,0)} \} }{ \sqrt{\det \{U_{(\pi,\pi),(\pi,-\pi)}\}}}
\end{equation}
for the path
\begin{equation}\label{path2}
{\bm k}=(\pi,-\pi) \rightarrow {\bm k}=(\pi,\pi).
\end{equation} 
We now form the product 
\begin{equation}
\Xi_{2D}= \Xi_{0} \Xi_{\pi},
\end{equation} 
in which case the arbitrariness in choosing the branch of the square root at the denominators becomes irrelevant because now we have a canonical way to chose the {\it same} branch for the square roots of $\det\{U_{(0,-\pi),(0,\pi)}\}$ and $\det\{U_{(\pi,-\pi),(\pi,\pi)}\}$. Indeed, the paths described in Eqs.~\ref{path1} and \ref{path2} can be deformed into each other without breaking the loops or leaving the Brillouing torus. Therefore, the Bloch Hamiltonians  $H(0,k_y)$ and $H(\pi,k_y)$ can be adiabatically connected without closing the energy gap, which means $U_{(0,-\pi),(0,\pi)}$ can be continuously evolved into $U_{(\pi,-\pi),(\pi,\pi)}$, and same can be said for their corresponding determinants. Therefore we have an effective way to make sure we choose the same branch of the square root for both determinants. It is totally irrelevant which branch we chose (as long is the same) because if we change the branch for both square roots, then a minus sign appears twice and nothing changes. The conclusion is that $\Xi_{2D}$ can be given a meaningful topological content, and different time-reversal invariant systems can be classified according to the corresponding value of $\Xi_{2D}$. The trivial insulator is contained in the class with $\Xi_{2D}=+1$ and the topologically non-trivial insulators are contained in the class with $\Xi_{2D}=-1$. We could have started the entire construction from paths oriented along the $k_x$ direction, but this would have led to the same topological invariant (see the argument by Roy in Ref.~\onlinecite{Roy2010nj}). 

Let us follow right away  with a non-trivial example. We chose to work with the Bernevig-Huges-Zhang model,\cite{Bernevig:2006hl} including the $Sz$-nonconserving term discussed in Ref.~\onlinecite{Yamakage2010xr}. The model is described by the Bloch Hamiltonians acting on the ${\bm C}^4$ space:
\begin{equation}\label{H2D0}
\begin{array}{c}
H_{\bm k}=\left ( 
\begin{array}{cc}
h({\bm k}) & \Gamma({\bm k}) \\
\Gamma({\bm k})^\dagger & h^*(-{\bm k})
\end{array}
\right ),
\end{array}
\end{equation}
where $h({\bm k})$$=$${\bm d}({\bm k})$$\cdot$${\bm \sigma}$, with ${\bm \sigma}$=$(\sigma_x,\sigma_y,\sigma_z)$ and:
\begin{equation}
{\bm d}= (A \sin k_x,A \sin k_y, \Delta-2B(2-\cos k_x - \cos k_y)).
\end{equation}
The $\Gamma$ term is given by:\cite{Yamakage2010xr}
\begin{equation}
\begin{array}{c}
\Gamma(k)=
i \Lambda \left (
\begin{array}{cc}
\sin k_x - i \sin k_y & 0 \\
0 & \sin k_x +i \sin k_y
\end{array}
\right ).
\end{array}
\end{equation}

As written above, the model is symmetric to time-reversal and to inversion symmetry operations. We include an additional term which will specifically break the inversion symmetry but leaves the time-reversal symmetry intact. To be as explicit as possible, let us mention that the action of the time-reversal operation $\theta=e^{i\pi S_y}K$ ($K$= complex conjugation) in ${\bm C}^4$ is:
 \begin{equation}\label{Theta}
 \theta \left (
 \begin{array}{c}
 a \\
 b  \\
 c \\
 d \\
 \end{array}
 \right)=\left (
 \begin{array}{cccc}
 0 & 0 & 1 & 0 \\
 0 & 0 & 0 & 1 \\
 -1 & 0 & 0 & 0 \\
 0 & -1 & 0 & 0
 \end{array}
 \right ) \left (
 \begin{array}{c}
 a^* \\
 b^*  \\
 c^* \\
 d^* \\
 \end{array}
 \right)
 \end{equation}
 The inversion operation is implemented by:
 \begin{equation}
P=\left (
 \begin{array}{cccc}
 1 & 0 & 0 & 0 \\
 0 & -1 & 0 & 0 \\
 0 & 0 & -1 & 0 \\
 0 & 0 & 0 & 1
 \end{array}
 \right )
 \end{equation} 
 The additional term to the Hamiltonian that we consider here is:
 \begin{equation}\label{H2D1}
 R\left (
 \begin{array}{cccc}
 0 & 0 & 0 & e^{ik_1+2ik_2} \\
 0 & 0 & -e^{-ik_1-2ik_2} & 0 \\
 0 & -e^{ik_1+2ik_2} & 0 & 0 \\
 e^{-ik_1-2ik_2} & 0 & 0 & 0
 \end{array}
 \right )
\end{equation}
The factor 2 in front of $k_2$ was chosen just to introduce an anisotropy. $R$ is the coupling constant.

\begin{figure*}
  \includegraphics[width=16cm]{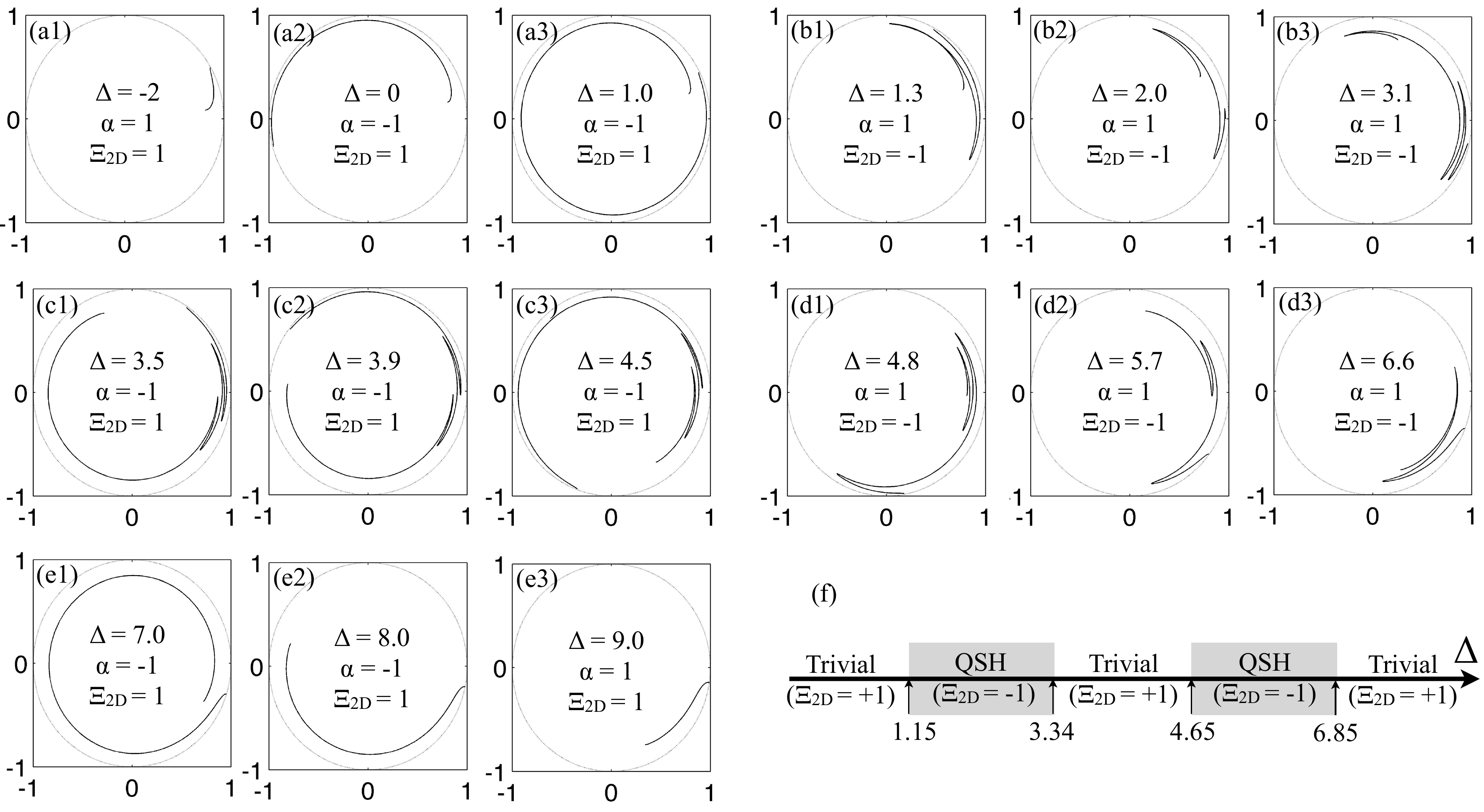}\\
  \caption{Results for the 2-dimensional model of Eqs.~\ref{H2D0} and \ref{H2D1}, with the parameters fixed at: $A=B=1$, $\Lambda=0.5$ and $R=2$. Each panel shows the path in the complex plane of $\det\{U_{(k_x,-\pi),(k_x,\pi)}\}$ as $k_x$ is varied from 0 to $\pi$, for different values of $\Delta$. The panels are grouped into bundles of 3 (for example d1, d2 and d3), and each such bundle samples a region of the phase diagram where the energy gap stays open. In each panel, one can read the value of $\Delta$, the correcting factor $\alpha$ and the value of the ${\bm Z}_2$ invariant $\Xi_{2D}$. Panel (f) shows the predicted phase diagram of the 2-dimensional  model. }
 \label{Fig1}
\end{figure*}

If $R=0$, $H_{\bm k}$ displays topological phases for $0<\Delta/B<4$ and $4$$<$$\Delta/B$$<$$8$, and trivial phases for $\Delta/B$$<$0 or $\Delta/B$$>$8.\cite{Yamakage2010xr,Prodan2011vy} The insulating gap closes when $\Delta/B$=0, 4 and 8. For a finite $R$, the phase diagram changes; the energy gap closes at 4 points and a new topologically trivial phase appears. Let us be explicit and fix some parameters, from now on, as follows: $A=B=1$, $\Lambda=0.5$ and $R=1$. Upon varying the parameter $\Delta$, we found that the energy gap closes at 1.15, 3.34, 4.65 and 6.85. By just taking into account the known phase diagram at $R=0$,\cite{Yamakage2010xr,Prodan2011vy} it is naturally to assume that the topological phases occur when $\Delta$ is in between 1.15 and 3.34, and in between 4.65 and 6.85. Topologically trivial phases are expected in rest (see Fig.~\ref{Fig1}f). But this we are going to check explicitly.

\begin{figure}
  \includegraphics[width=7cm]{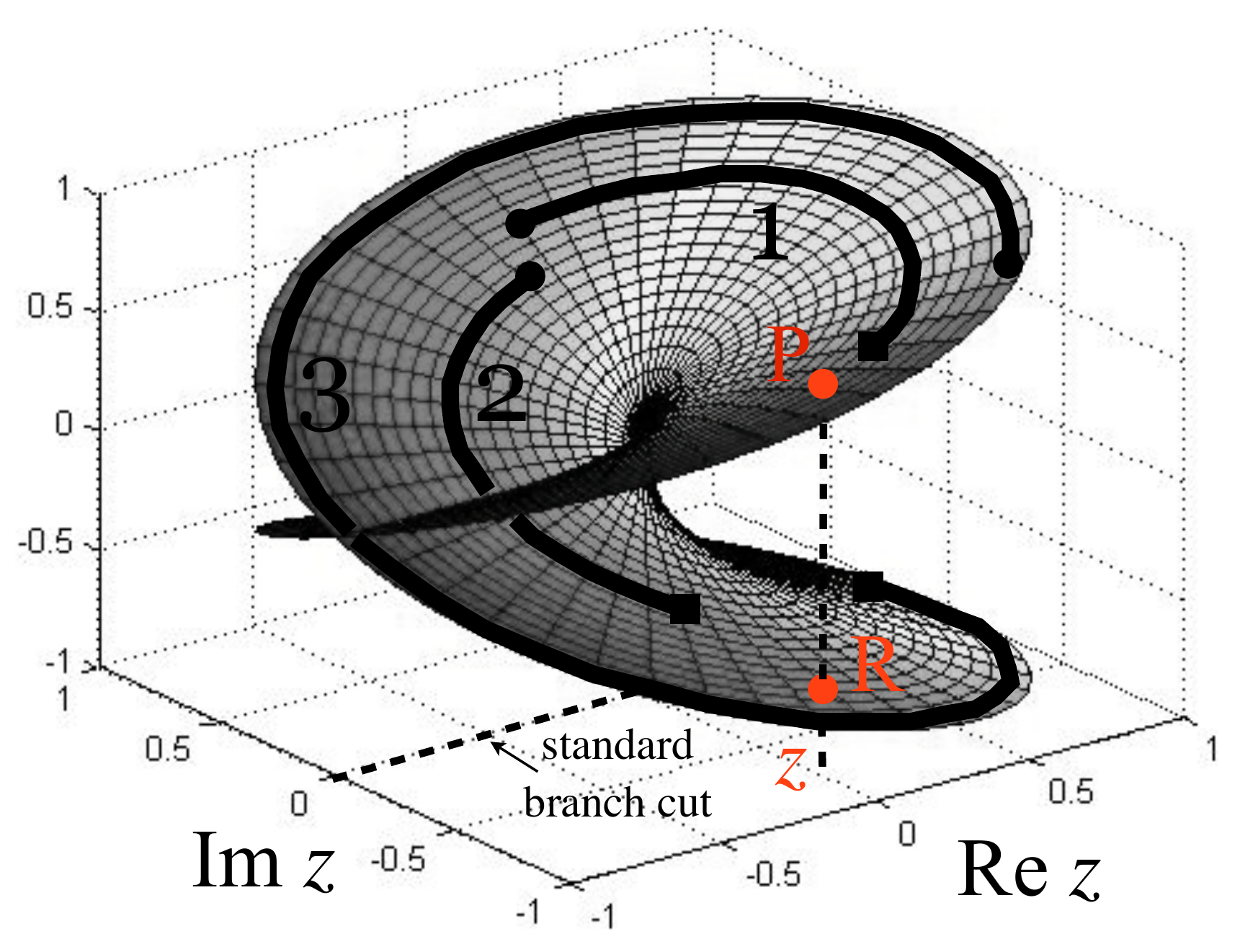}\\
  \caption{(Color online) The complex function$\sqrt{z}$ is multivalued and its proper representation is on a Riemann surface, shown in this figure. The Riemann surface consists of top and bottom sheets, which are connected along the segment $(-\infty, 0]$ of the real axis (the cut of the Riemann surface into sheets is not unique, but this is the standard cut adopted by most scientific softwares). Given a point $z$ in the complex plane, we can assign to it two points (see P and R in the diagram) on the Riemann surface. Most scientific software will automatically assign point P and compute the square root of $z$ accordingly. If point R would have been assigned, then the square root would differ by exactly a factor $-1$. When computing the square root of a $z$ that is continuously varied, one can easily map the position of $z$ on the Riemann surface and therefore correct the value computed by the software. As examples, we considered three paths for $z$, starting at the solid circle and ending at the solid square. The square root at the end of each path must be corrected by: $\alpha = +1$ (no correction) for path 1 and $\alpha=-1$ for paths 2 and 3.}
 \label{Fig2}
 \end{figure}

The first part of the calculation relates to the square root of the determinants of the monodromies. For this we have considered paths along the $k_y$ direction:
\begin{equation}
{\bm k}=(k_x,-\pi) \rightarrow {\bm k}=(k_x,\pi),
\end{equation}
 for which we computed the monodromy $U_{(k_x,-\pi),(k_x,\pi)}$, by straight implementation of the Eq.~\ref{Solution}, using 1000 discretization points. We will be more specific about this part of the calculation shortly. We then plotted in the complex plane the value of the determinant of the monodromies as function of $k_x$, when $k_x$ varied from $0$ to $\pi$ (here we used again 1000 discretization points). Fig.~\ref{Fig1} shows the plots for different values of $\Delta$. When taking the square root of the determinant, what we must have in mind is the Riemann surface of the complex function $\sqrt{z}$, shown in Fig.~\ref{Fig2}. Most of the available softwares, when given a complex number $z$ in the plane, it will automatically place $z$ on the top sheet of the Riemann surface. As explained above, we can chose any branch of the square root for the determinant at $k_x=0$, but after that we must be consistent with this choice when we compute the square root of the determinant at $k_x=\pi$. So we will always place the determinant at $k_x=0$ on the top sheet of the Riemann surface. Then, by following the evolution of the determinant of the monodromy as $k_x$ is varied from $0$ to $\pi$, we will be able to tell exactly where this determinant is located on the Riemann surface. If the determinant ends up on the top sheet, we don't need any correction, but if it ends up on the lower sheet, we must correct the output from the software by multiplying the square root by a correction factor $\alpha=-1$. In Fig.~\ref{Fig2} we chose several situations and explain in detail how $\alpha$ works. To summarize, in the actual calculation we let the software (in this case MATLAB) to compute the square root of the determinants and afterwards we corrected the square root of the determinant at $k_x=\pi$ by the sign factor $\alpha$, which was determined from the inspection of the graphs in Fig.~\ref{Fig1} and using the prescription given in Fig.~\ref{Fig2}. While here we chose to use a visual inspection to determine $\alpha$, and this was mainly to show the reader how things work, it is important to notice that $\alpha$ can be determined in an automated fashion, without any visual inspection. This observation is important for the implementation of the formalism in the first principle codes.\cite{Soluyanov2011gy} 
 
The second step of the calculation consist of evaluating the Pfaffians and the monodromies at the nominators in Eqs.~\ref{Xi1} and \ref{Xi2}. These were numerically evaluated in the following way. The line between $k_y$=0 and $k_y$=$\pi$ (this time assuming $k_x$ to be either 0 or $\pi$) was discretized using 1000 points and $H_k$ was diagonalized at all these $k_y$-points. This provided us with four eigenvectors $\psi_i(k_y)$, $i=1,\ldots,4$, sorted according to their eigenvalues. Being part of the ${\bm C}^4$ space, the eigenvectors are represented as 4-component column matrices. Note that the diagonalization procedure gives random phases for the eigenvectors, but this is irrelevant when we form the projector onto the first two egivenvectors:
\begin{equation}
P_{k_y} = |\psi_1(k_y) \rangle \langle \psi_1(k_y)| + |\psi_2(k_y) \rangle \langle \psi_2(k_y)|.
\end{equation}
The projectors were represented as 4$\times$4 matrices. As $k_y$ was progressing from 0 to $\pi$, we have continuously updated the monodromy matrix: $U$$\rightarrow$$P_{k_y} U$, starting initially from $U$=$P_{k_y=0}$. After the monodromy was computed, we used the bases $\{\psi_{1}(0),\psi_{2}(0)\}$ and $\{\psi_{1}(\pi),\psi_{2}(\pi)\}$ for the occupied spaces at $k_y=0$ and $\pi$ to compute the 2$\times$2 matrix $\hat{U}$.

The Pfaffians at $k_y=0$ and $\pi$ are simply equal to $\langle \psi_1(0)|\theta|\psi_2(0)\rangle$ and $\langle \psi_1(\pi)|\theta|\psi_2(\pi)\rangle$, respectively, and these matrix elements were easily computed using the action shown in Eq.~\ref{Theta}.  One important note here is that the Pfaffians have to be computed using the same bases (including the phases!) as the ones used to compute $\hat{U}$. One can verified how converged the monodromies are, by examining the absolute values of their determinants. In general, these values will be less than 1, but converge towards this ideal value of 1 as more points are added to the discretization of the paths.

 We now return to Fig.~\ref{Fig1} and discuss the results. We picked three $\Delta$ values in each region of the phase diagram, so that we have values close to the end points where the gap closes and values far away from these points. The actual values are shown in the middle of each panel. Each panel shows, in the complex plane, the calculated value for $\det\{U_{(k_x,-\pi),(k_x,\pi)}\}$, as $k_x$ was varied from $0$ to $\pi$. This determinant is always on the unit circle, but just for a better representation, we have artificially shifted its value inside the unit circle (by multiplying with the function $e^{-\frac{0.2k_x}{\pi}}$), so that we can follow its intricate behavior. Given these curves, and assuming that the determinant at $k_x=0$ was on the upper sheet of the Riemann surface of the square root, we can easily determine the position of the determinant at $k_x=\pi$ on the Riemann surface of $\sqrt{z}$, and therefore the value of $\alpha$ (see the discussion in Fig.~\ref{Fig2}). We have placed these values directly inside the panels, together with the value of the ${\bm Z}_2$ invariant. Besides these calculations, we have performed calculations with a much more refined sampling of $\Delta$, confirming the phase diagram shown in in panel (f) of Fig.~\ref{Fig1}.

 \section{The ${\bm Z}_2$ topological invariants in 3-dimensions}

In three dimensions, we can use different pairs of time-reversal invariant paths and construct weak ${\bm Z}_2$ invariants first. Let us consider the following explicit pairs: 
\begin{equation}\label{pair1}
\mbox{pair 1:} \left \{\begin{array}{l}
 \ {\bm k}=(0,0,-\pi) \rightarrow {\bm k}=(0,0,\pi) \\
 \ {\bm k}=(0,\pi,-\pi) \rightarrow {\bm k}=(0,\pi,\pi)
\end{array}\right .
\end{equation}
and
\begin{equation}\label{pair2}
\mbox{pair 2:} \left \{\begin{array}{l}
  {\bm k}=(\pi,0,-\pi) \rightarrow {\bm k}=(\pi,0,\pi) \\
 \ {\bm k}=(\pi,\pi,-\pi) \rightarrow {\bm k}=(\pi,\pi,\pi),  
\end{array}\right .
\end{equation}
for which we construct the corresponding 2-dimensional (weak) ${\bm Z}_2$ invariants, $\Xi_{2D}$ and $\Xi'_{2D}$, as described in the previous section. All we have to check, and this is obvious, is that the paths in each pairs can be deformed into each other continuously without leaving the Brillouin torus (in fact, all four paths can be deformed into each other). The strong invariant is given by their product: 
\begin{equation}\label{Xi3D}
\Xi_{3D}=\Xi_{2D} \Xi'_{2D}.
\end{equation}
We can start the construction from different pairs of paths, but at the end we can generate at most 3 independent weak invariants plus the unique strong invariant, a fact that can be shown by using a fairly general method introduced by Roy.\cite{Roy2010nj}

Let us again follow with an example. We chose to work with the model Hamiltonians reported in Ref.~\onlinecite{LiuPRB2010xf}. We will tune the parameters for Bi$_2$Se$_3$, following Ref.~\onlinecite{Qi2010hg} (see Eq.~31 and Table II). Explicitly, we considered the Bloch Hamiltonians:
\begin{equation}\label{H3D1}
H_{\bm k}=\left (
\begin{array}{cccc}
M & A_1 & 0 & A_2 \\
A_1 & -M & A_2 & 0 \\
0 & A_2^* & M & -A_1 \\
A_2^* & 0 & -A_1 & -M
\end{array}
\right ),
\end{equation}
with (eV units are assumed):
\begin{equation}\label{H3D2}
\begin{array}{c}
M=13.72\cos k_z + 89\cos \sqrt{k^2-k_z^2} + \Delta-102.72 \medskip \\
A_1=2.26 \sin k_z, \ A_2=3.33(\sin k_x -i \sin k_y). 
\end{array}
\end{equation}
The parameter $\Delta$ was allowed to vary. 

The action of the time-reversal operation in ${\bm C}^4$ is the same as described in Eq.~\ref{Theta} and the inversion operation is implemented by:
 \begin{equation}
P=\left (
 \begin{array}{cccc}
 1 & 0 & 0 & 0 \\
 0 & -1 & 0 & 0 \\
 0 & 0 & 1 & 0 \\
 0 & 0 & 0 & -1
 \end{array}
 \right ).
 \end{equation}  
 As written above, the model has both time-reversal and inversion symmetries. Therefore, we introduce an additional term in the Hamiltonian, which breaks the inversion symmetry:
  \begin{equation}\label{H3D3}
 R\left (
 \begin{array}{cccc}
 0 & 0 & 0 & e^{-ik_3} \\
 0 & 0 & -e^{ik_3} & 0 \\
 0 & -e^{-ik_3} & 0 & 0 \\
 e^{ik_3} & 0 & 0 & 0
 \end{array}
 \right )
\end{equation}

\begin{figure*}
  \includegraphics[width=16cm]{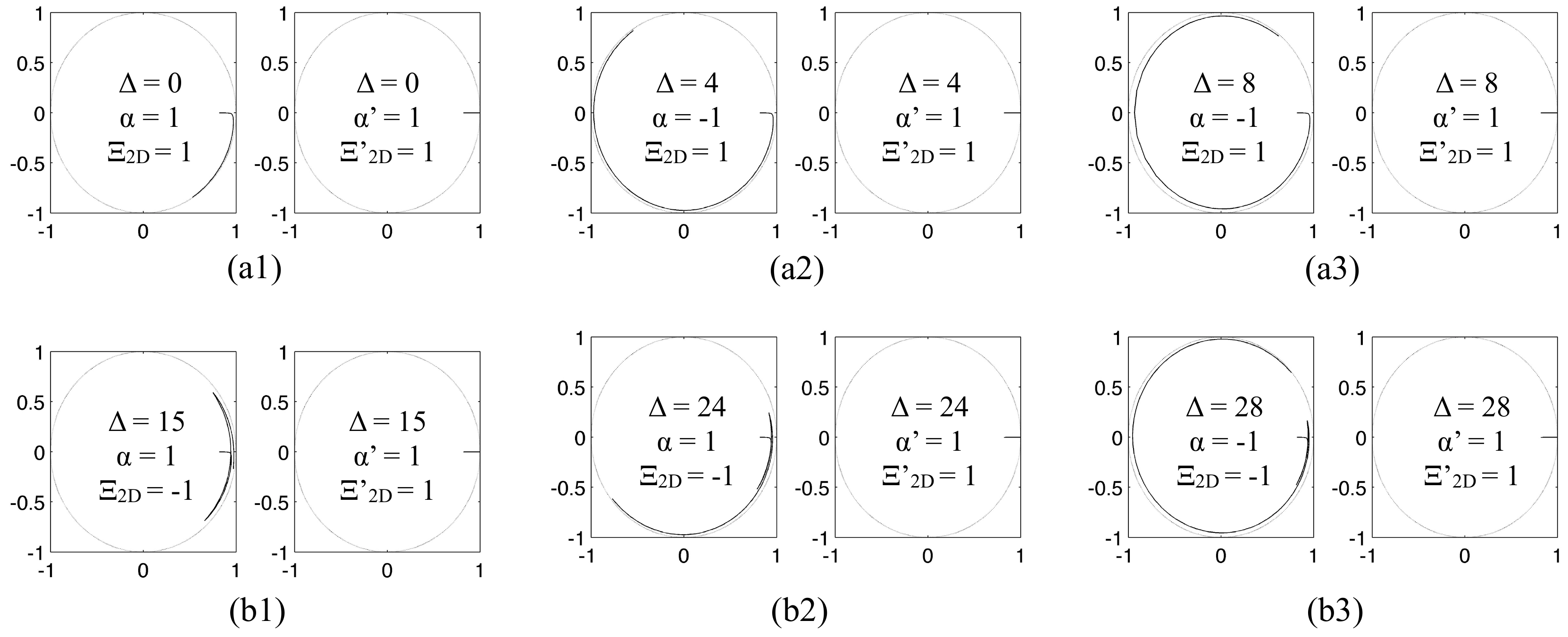}\\
  \caption{Results for the 3-dimensional model of Eqs.~\ref{H3D1}, \ref{H3D2} and \ref{H3D3}, with $R=2$. Each panel contains two plots, showing the path in the complex plane of $\det\{U_{(0,k_y,-\pi),(0,k_y,\pi)}\}$ (left plot), and of $\det\{U_{(\pi,k_y,-\pi),(\pi,k_y,\pi)}\}$ (right plot), both as functions of $k_y$, which is varied from 0 to $\pi$. The values of $\Delta$ and of the resulting $\alpha$ and $\Xi_{2D}$ are also shown. The strong invariant $\Xi_{3D}$ is computed to be +1 for panels (a) and -1 for panels (b). The determinants at $k_x=\pi$ seem to be pinned at +1, which is a peculiarity of the model.}
 \label{Fig3}
\end{figure*}
 
Even for $R=0$, when inversion symmetry is present, the model displays a fairly complex phase diagram as function of $\Delta$. The energy gap closes at $\Delta=0$, 27.44, 116.44, 140.14, 178, 205.44, and the model displays QSH (${\bm Z}_2=-1$) phases inside the intervals: $(0,27.44)$, $(116.44,140.14)$ and $(178,205.44)$. This can be verified directly by computing the parities at the time-reversal invariant $k$-points. We have verified that our formula for the strong ${\bm Z}_2$ invariant, given in Eq.~\ref{Xi3D}, gives the same results. We will not present these calculations here and instead we will present in detail the case when the inversion symmetry is absent. For this we chose $R=2$, in which case the phase diagram as function of $\Delta$ is qualitatively changed, by the emergence of few metallic phases. We will focus only on the lower part of the diagram, where direct bands structure calculations show that the energy gap closes at $\Delta=10$ and that it remains closed until $\Delta = 13$. After that the gap stays opened until $\Delta=30$, when the gap closes and remains closed when $\Delta$ is further increased. Additional phases emerge after that, but will not be discussed here. The numerical results are shown in Fig.~\ref{Fig3}, where the two insulating phases mentioned above are sampled in 3 points. The calculation shows that the insulating phase below $\Delta=10$ is trivial, while the one between $\Delta=13$ and $\Delta=30$ is topologically non-trivial. 

\section{Extension to continuum models}

Let us consider a periodic crystal described by a Hamiltonian (${\bm R}=$ a lattice vector):
\begin{equation}
H=-{\bm \nabla}^2+\hat{V}({\bm r}), \ \ \hat{V}({\bm r+\bm R})=\hat{V}({\bm r})
\end{equation}
acting on the Hilbert space ${\cal H}$ of 2-component spinors that are square integrable. The Hamiltonian is assumed to commute with the time-reversal operation $e^{i\pi S_y}K$ ($K$ = complex conjugation).

We now consider the Bloch decomposition, given by the isometry:
\begin{equation}
\begin{array}{c}
u:{\cal H} \rightarrow \oplus_{\bm k} \tilde{{\cal H}}, \ \ u\Psi=\oplus_{\bm k} \Psi_{\bm k}, \medskip \\
 \ \ \Psi_{\bm k}({\bm r})=\sum_{\bm R} e^{-i{\bm k}\cdot{\bm R}}\Psi({\bm r}+{\bm R}),
\end{array}
\end{equation}
where ${\cal H}$ is the original Hilbert space and $\tilde{{\cal H}}$ represents the space of square integrable spinors defined over only one unit cell. Under this isometry, we have:
\begin{equation}
uHu^{-1}=\oplus_{\bm k} H_{\bm k},
\end{equation}
where $H_{\bm k}$ is given by $-{\bm \nabla}^2+\hat{V}({\bm r})$ but this time defined only over one unit cell and with the Bloch boundary conditions (the prime indicates the derivative):
\begin{equation}\label{BlochBC}
\begin{array}{c}
\Psi_{\bm k}({\bm r}+{\bm R})=e^{i{\bm k} \cdot {\bm R}}\Psi_{\bm k}({\bm r}) \medskip \\
\Psi'_{\bm k}({\bm r}+{\bm R})=e^{i{\bm k} \cdot {\bm R}}\Psi'_{\bm k}({\bm r})
\end{array}
\end{equation}
whenever ${\bm r}$ and ${\bm r}+{\bm R}$ are on the boundaries of the unit cell. The time-reversal operation satisfies:
\begin{equation}
\theta H_{\bm k} \theta^{-1} = H_{-{\bm k}},
\end{equation}
so at this point there is no practical difference between the continuum model and the tight-binding models discussed in the previous sections. 

The projectors $P_{\bm k}$ of the Hamiltonians $H_{\bm k}$ onto the states below a given Fermi level are routinely computed by the first principle codes. Hence, the monodromies, the $\Xi_{2D}$ and $\Xi_{3D}$ invariants can be computed in a straightforward fashion. We are currently working on implementing the whole construction in our first principle code but, unfortunately, we cannot show any concrete results at this time.   

\section{Conclusions}

Using a monodromy technique, we proposed new formulations of the ${\bm Z}_2$ invariants for topological insulators with time-reversal symmetry. The formulations are manifestly gauge independent and we argued that they can be effortlessly integrated in the tight-binding as well as first principle simulations. Test calculations confirmed a full agreement between the new formulations and the already established ones. We hope that the expressions of the ${\bm Z}_2$ invariants given in this work will help the scientists searching for novel non centro-symmetric 3 dimensional topological insulators. We are currently integrating the new formulations of the ${\bm Z}_2$ invariants in our first principle electronic structure codes as an automated post-processing routines. We hope that other electronic structure practitioners will follow our example. One other hope of ours is that this gauge independent formulations will lead to more effective and transparent real space formulations of the ${\bm Z}_2$ invariants, absolutely necessary for understanding the disorder effects in time-reversal invariant topological insulators.\cite{ProdanJPhysA2010xk} 

At the end, let us comment about the monodromy technique. It definitely helped us avoid complex calculations, as the present results followed entirely from the group property and the behavior under time-reversal of the monodromy. So far, the monodromy technique has been applied to inversion symmetric insulators,\cite{Hughes2010gh} to filamentary structures with inversion symmetry supporting topological phonon modes,\cite{BergPRE2011vy}  and to time-reversal symmetric insulators. It is very likely that other point symmetries could be handled in a similar fashion, which is a future direction that we are currently exploring.

\begin{acknowledgments} 
This research was supported by a Cottrell award from the Research Corporation for Science Advancement and by the office of the Provost of Yeshiva University. We want to thank David Vanderbilt for spotting a flaw in our original argument and for extremely useful discussions about the subject.
\end{acknowledgments}


\begin{thebibliography}{10}%
\makeatletter
\providecommand \@ifxundefined [1]{%
 \ifx #1\undefined \expandafter \@firstoftwo
 \else \expandafter \@secondoftwo
\fi
}%
\providecommand \@ifnum [1]{%
 \ifnum #1\expandafter \@firstoftwo
 \else \expandafter \@secondoftwo
\fi
}%
\providecommand \enquote [1]{``#1''}%
\providecommand \bibnamefont  [1]{#1}%
\providecommand \bibfnamefont [1]{#1}%
\providecommand \citenamefont [1]{#1}%
\providecommand\href[0]{\@sanitize\@href}%
\providecommand\@href[1]{\endgroup\@@startlink{#1}\endgroup\@@href}%
\providecommand\@@href[1]{#1\@@endlink}%
\providecommand \@sanitize [0]{\begingroup\catcode`\&12\catcode`\#12\relax}%
\@ifxundefined \pdfoutput {\@firstoftwo}{%
 \@ifnum{\z@=\pdfoutput}{\@firstoftwo}{\@secondoftwo}%
}{%
 \providecommand\@@startlink[1]{\leavevmode}%
 \providecommand\@@endlink[0]{}%
}{%
 \providecommand\@@startlink[1]{%
  \leavevmode
  \pdfstartlink
   attr{/Border[0 0 1 ]/H/I/C[0 1 1]}%
   user{/Subtype/Link/A<</Type/Action/S/URI/URI(#1)>>}%
  \relax
 }%
 \providecommand\@@endlink[0]{\pdfendlink}%
}%
\providecommand \url  [0]{\begingroup\@sanitize \@url }%
\providecommand \@url [1]{\endgroup\@href {#1}{\urlprefix}}%
\providecommand \urlprefix [0]{URL }%
\providecommand \Eprint[0]{\href }%
\@ifxundefined \urlstyle {%
  \providecommand \doi [1]{doi:\discretionary{}{}{}#1}%
}{%
  \providecommand \doi [0]{doi:\discretionary{}{}{}\begingroup
  \urlstyle{rm}\Url }%
}%
\providecommand \doibase [0]{http://dx.doi.org/}%
\providecommand \Doi[1]{\href{\doibase#1}}%
\providecommand \bibAnnote [3]{%
  \BibitemShut{#1}%
  \begin{quotation}\noindent
    \textsc{Key:}\ #2\\\textsc{Annotation:}\ #3%
  \end{quotation}%
}%
\providecommand \bibAnnoteFile [2]{%
  \IfFileExists{#2}{\bibAnnote {#1} {#2} {\input{#2}}}{}%
}%
\providecommand \typeout [0]{\immediate \write \m@ne }%
\providecommand \selectlanguage [0]{\@gobble}%
\providecommand \bibinfo [0]{\@secondoftwo}%
\providecommand \bibfield [0]{\@secondoftwo}%
\providecommand \translation [1]{[#1]}%
\providecommand \BibitemOpen[0]{}%
\providecommand \bibitemStop [0]{}%
\providecommand \bibitemNoStop [0]{.\EOS\space}%
\providecommand \EOS [0]{\spacefactor3000\relax}%
\providecommand \BibitemShut [1]{\csname bibitem#1\endcsname}%
\bibitem{HALDANE:1988rh}%
  \BibitemOpen
  \bibfield{author}{%
  \bibinfo {author} {\bibfnamefont{F.~D.~M.}\ \bibnamefont{Haldane}},\ }%
  \bibfield{journal}{%
  \bibinfo {journal} {Phys. Rev. Lett.}\ }%
  \textbf{\bibinfo {volume} {61}},\ \bibinfo {pages} {2015} (\bibinfo {year}
  {1988})%
  \bibAnnoteFile{NoStop}{HALDANE:1988rh}%
\bibitem{Kane:2005np}%
  \BibitemOpen
  \bibfield{author}{%
  \bibinfo {author} {\bibfnamefont{C.~L.}\ \bibnamefont{Kane}}\ and\ \bibinfo
  {author} {\bibfnamefont{E.~J.}\ \bibnamefont{Mele}},\ }%
  \bibfield{journal}{%
  \bibinfo {journal} {Phys. Rev. Lett.}\ }%
  \textbf{\bibinfo {volume} {95}},\ \bibinfo {pages} {226801} (\bibinfo {year}
  {2005})%
  \bibAnnoteFile{NoStop}{Kane:2005np}%
\bibitem{Kane:2005zw}%
  \BibitemOpen
  \bibfield{author}{%
  \bibinfo {author} {\bibfnamefont{C.~L.}\ \bibnamefont{Kane}}\ and\ \bibinfo
  {author} {\bibfnamefont{E.~J.}\ \bibnamefont{Mele}},\ }%
  \bibfield{journal}{%
  \bibinfo {journal} {Phys. Rev. Lett.}\ }%
  \textbf{\bibinfo {volume} {95}},\ \bibinfo {pages} {146802} (\bibinfo {year}
  {2005})%
  \bibAnnoteFile{NoStop}{Kane:2005zw}%
\bibitem{Prodan:2009lo}%
  \BibitemOpen
  \bibfield{author}{%
  \bibinfo {author} {\bibfnamefont{E.}~\bibnamefont{Prodan}},\ }%
  \bibfield{journal}{%
  \bibinfo {journal} {J. Math. Phys.}\ }%
  \textbf{\bibinfo {volume} {50}},\ \bibinfo {pages} {083517} (\bibinfo {year}
  {2009})%
  \bibAnnoteFile{NoStop}{Prodan:2009lo}%
\bibitem{Prodan:2009mi}%
  \BibitemOpen
  \bibfield{author}{%
  \bibinfo {author} {\bibfnamefont{E.}~\bibnamefont{Prodan}},\ }%
  \bibfield{journal}{%
  \bibinfo {journal} {J. Phys. A: Math. Theor.}\ }%
  \textbf{\bibinfo {volume} {42}},\ \bibinfo {pages} {082001} (\bibinfo {year}
  {2009})%
  \bibAnnoteFile{NoStop}{Prodan:2009mi}%
\bibitem{Bernevig:2006hl}%
  \BibitemOpen
  \bibfield{author}{%
  \bibinfo {author} {\bibfnamefont{B.~A.}\ \bibnamefont{Bernevig}}, \bibinfo
  {author} {\bibfnamefont{T.~L.}\ \bibnamefont{Hughes}},\ and\ \bibinfo
  {author} {\bibfnamefont{S.-C.}\ \bibnamefont{Zhang}},\ }%
  \bibfield{journal}{%
  \bibinfo {journal} {Science}\ }%
  \textbf{\bibinfo {volume} {314}},\ \bibinfo {pages} {1757} (\bibinfo {year}
  {2006})%
  \bibAnnoteFile{NoStop}{Bernevig:2006hl}%
\bibitem{Koenig:2007ko}%
  \BibitemOpen
  \bibfield{author}{%
  \bibinfo {author} {\bibfnamefont{M.}~\bibnamefont{Koenig}}, \bibinfo {author}
  {\bibfnamefont{S.}~\bibnamefont{Wiedmann}}, \bibinfo {author}
  {\bibfnamefont{C.}~\bibnamefont{Bruene}}, \bibinfo {author}
  {\bibfnamefont{A.}~\bibnamefont{Roth}}, \bibinfo {author}
  {\bibfnamefont{H.}~\bibnamefont{Buhmann}}, \bibinfo {author}
  {\bibfnamefont{L.~W.}\ \bibnamefont{Molenkamp}}, \bibinfo {author}
  {\bibfnamefont{X.-L.}\ \bibnamefont{Qi}},\ and\ \bibinfo {author}
  {\bibfnamefont{S.-C.}\ \bibnamefont{Zhang}},\ }%
  \bibfield{journal}{%
  \bibinfo {journal} {Science}\ }%
  \textbf{\bibinfo {volume} {318}},\ \bibinfo {pages} {766} (\bibinfo {year}
  {2007})%
  \bibAnnoteFile{NoStop}{Koenig:2007ko}%
\bibitem{Fu:2007vs}%
  \BibitemOpen
  \bibfield{author}{%
  \bibinfo {author} {\bibfnamefont{L.}~\bibnamefont{Fu}}\ and\ \bibinfo
  {author} {\bibfnamefont{C.~L.}\ \bibnamefont{Kane}},\ }%
  \bibfield{journal}{%
  \bibinfo {journal} {Phys. Rev. B}\ }%
  \textbf{\bibinfo {volume} {76}},\ \bibinfo {pages} {045302} (\bibinfo {year}
  {2007})%
  \bibAnnoteFile{NoStop}{Fu:2007vs}%
\bibitem{Hsieh:2008vm}%
  \BibitemOpen
  \bibfield{author}{%
  \bibinfo {author} {\bibfnamefont{D.}~\bibnamefont{Hsieh}}, \bibinfo {author}
  {\bibfnamefont{D.}~\bibnamefont{Qian}}, \bibinfo {author}
  {\bibfnamefont{L.}~\bibnamefont{Wray}}, \bibinfo {author}
  {\bibfnamefont{Y.}~\bibnamefont{Xia}}, \bibinfo {author}
  {\bibfnamefont{Y.~S.}\ \bibnamefont{Hor}}, \bibinfo {author}
  {\bibfnamefont{R.~J.}\ \bibnamefont{Cava}},\ and\ \bibinfo {author}
  {\bibfnamefont{M.~Z.}\ \bibnamefont{Hasan}},\ }%
  \bibfield{journal}{%
  \bibinfo {journal} {Nature}\ }%
  \textbf{\bibinfo {volume} {452}},\ \bibinfo {pages} {970} (\bibinfo {year}
  {2008})%
  \bibAnnoteFile{NoStop}{Hsieh:2008vm}%
\bibitem{Moore:2007ew}%
  \BibitemOpen
  \bibfield{author}{%
  \bibinfo {author} {\bibfnamefont{J.~E.}\ \bibnamefont{Moore}}\ and\ \bibinfo
  {author} {\bibfnamefont{L.}~\bibnamefont{Balents}},\ }%
  \bibfield{journal}{%
  \bibinfo {journal} {Phys. Rev. B}\ }%
  \textbf{\bibinfo {volume} {75}},\ \bibinfo {pages} {121306} (\bibinfo {year}
  {2007})%
  \bibAnnoteFile{NoStop}{Moore:2007ew}%
\bibitem{Loring2010vy}%
  \BibitemOpen
  \bibfield{author}{%
  \bibinfo {author} {\bibfnamefont{T.~A.}\ \bibnamefont{Loring}}\ and\ \bibinfo
  {author} {\bibfnamefont{M.~B.}\ \bibnamefont{Hastings}},\ }%
  \bibfield{journal}{%
  \bibinfo {journal} {Europhys. Lett.}\ }%
  \textbf{\bibinfo {volume} {92}},\ \bibinfo {pages} {67004} (\bibinfo {year}
  {2010})%
  \bibAnnoteFile{NoStop}{Loring2010vy}%
\bibitem{Panati:2007sy}%
  \BibitemOpen
  \bibfield{author}{%
  \bibinfo {author} {\bibfnamefont{G.}~\bibnamefont{Panati}},\ }%
  \bibfield{journal}{%
  \bibinfo {journal} {Ann. Henri Poincare}\ }%
  \textbf{\bibinfo {volume} {8}},\ \bibinfo {pages} {995} (\bibinfo {year}
  {2007})%
  \bibAnnoteFile{NoStop}{Panati:2007sy}%
\bibitem{Fu:2006ka}%
  \BibitemOpen
  \bibfield{author}{%
  \bibinfo {author} {\bibfnamefont{L.}~\bibnamefont{Fu}}\ and\ \bibinfo
  {author} {\bibfnamefont{C.~L.}\ \bibnamefont{Kane}},\ }%
  \bibfield{journal}{%
  \bibinfo {journal} {Phys. Rev. B}\ }%
  \textbf{\bibinfo {volume} {74}},\ \bibinfo {pages} {195312} (\bibinfo {year}
  {2006})%
  \bibAnnoteFile{NoStop}{Fu:2006ka}%
\bibitem{Fu:2007ti}%
  \BibitemOpen
  \bibfield{author}{%
  \bibinfo {author} {\bibfnamefont{L.}~\bibnamefont{Fu}}, \bibinfo {author}
  {\bibfnamefont{C.~L.}\ \bibnamefont{Kane}},\ and\ \bibinfo {author}
  {\bibfnamefont{E.~J.}\ \bibnamefont{Mele}},\ }%
  \bibfield{journal}{%
  \bibinfo {journal} {Phys. Rev. Lett.}\ }%
  \textbf{\bibinfo {volume} {98}},\ \bibinfo {pages} {106803} (\bibinfo {year}
  {2007})%
  \bibAnnoteFile{NoStop}{Fu:2007ti}%
\bibitem{FukuiJPSJ2005gu}%
  \BibitemOpen
  \bibfield{author}{%
  \bibinfo {author} {\bibfnamefont{T.}~\bibnamefont{Fukui}}, \bibinfo {author}
  {\bibfnamefont{Y.}~\bibnamefont{Hatsugai}},\ and\ \bibinfo {author}
  {\bibfnamefont{H.}~\bibnamefont{Suzuki}},\ }%
  \bibfield{journal}{%
  \bibinfo {journal} {J. Phys. Soc. of Japan}\ }%
  \textbf{\bibinfo {volume} {74}},\ \bibinfo {pages} {1674} (\bibinfo {year}
  {2005})%
  \bibAnnoteFile{NoStop}{FukuiJPSJ2005gu}%
\bibitem{FukuiJPSJ2007ny}%
  \BibitemOpen
  \bibfield{author}{%
  \bibinfo {author} {\bibfnamefont{F.}~\bibnamefont{T}}\ and\ \bibinfo {author}
  {\bibfnamefont{H.}~\bibnamefont{Y}},\ }%
  \bibfield{journal}{%
  \bibinfo {journal} {J. Phys. Soc. of Japan}\ }%
  \textbf{\bibinfo {volume} {76}},\ \bibinfo {pages} {145209} (\bibinfo {year}
  {2007})%
  \bibAnnoteFile{NoStop}{FukuiJPSJ2007ny}%
\bibitem{Essin:2007ij}%
  \BibitemOpen
  \bibfield{author}{%
  \bibinfo {author} {\bibfnamefont{A.~M.}\ \bibnamefont{Essin}}\ and\ \bibinfo
  {author} {\bibfnamefont{J.~E.}\ \bibnamefont{Moore}},\ }%
  \bibfield{journal}{%
  \bibinfo {journal} {{Phys. Rev. B}}\ }%
  \textbf{\bibinfo {volume} {{76}}},\ \bibinfo {pages} {{165307}} (\bibinfo
  {year} {{2007}})%
  \bibAnnoteFile{NoStop}{Essin:2007ij}%
\bibitem{XiaPRL2010ni}%
  \BibitemOpen
  \bibfield{author}{%
  \bibinfo {author} {\bibfnamefont{D.}~\bibnamefont{Xiao}}, \bibinfo {author}
  {\bibfnamefont{Y.~G.}\ \bibnamefont{Yao}}, \bibinfo {author}
  {\bibfnamefont{W.~X.}\ \bibnamefont{Feng}}, \bibinfo {author}
  {\bibfnamefont{J.}~\bibnamefont{Wen}}, \bibinfo {author}
  {\bibfnamefont{W.~G.}\ \bibnamefont{Zhu}}, \bibinfo {author}
  {\bibfnamefont{X.~Q.}\ \bibnamefont{Chen}}, \bibinfo {author}
  {\bibfnamefont{G.~M.}\ \bibnamefont{Stocks}},\ and\ \bibinfo {author}
  {\bibfnamefont{Z.~Y.}\ \bibnamefont{Zhang}},\ }%
  \bibfield{journal}{%
  \bibinfo {journal} {Phys. Rev. Lett.}\ }%
  \textbf{\bibinfo {volume} {105}},\ \bibinfo {pages} {096404} (\bibinfo {year}
  {2010})%
  \bibAnnoteFile{NoStop}{XiaPRL2010ni}%
\bibitem{FengPRB2010ry}%
  \BibitemOpen
  \bibfield{author}{%
  \bibinfo {author} {\bibfnamefont{W.~X.~F.}\ \bibnamefont{WX}}, \bibinfo
  {author} {\bibfnamefont{D.}~\bibnamefont{Xiao}}, \bibinfo {author}
  {\bibfnamefont{Y.}~\bibnamefont{Zhang}},\ and\ \bibinfo {author}
  {\bibfnamefont{Y.~G.}\ \bibnamefont{Yao}},\ }%
  \bibfield{journal}{%
  \bibinfo {journal} {Phys. Rev. B}\ }%
  \textbf{\bibinfo {volume} {82}},\ \bibinfo {pages} {235121} (\bibinfo {year}
  {2010})%
  \bibAnnoteFile{NoStop}{FengPRB2010ry}%
\bibitem{SoluyanovPRB2011bt}%
  \BibitemOpen
  \bibfield{author}{%
  \bibinfo {author} {\bibfnamefont{A.~A.}\ \bibnamefont{Soluyanov}}\ and\
  \bibinfo {author} {\bibfnamefont{D.}~\bibnamefont{Vanderbilt}},\ }%
  \bibfield{journal}{%
  \bibinfo {journal} {Phys. Rev. B}\ }%
  \textbf{\bibinfo {volume} {83}},\ \bibinfo {pages} {035108} (\bibinfo {year}
  {2011})%
  \bibAnnoteFile{NoStop}{SoluyanovPRB2011bt}%
\bibitem{FengPRL2010gu}%
  \BibitemOpen
  \bibfield{author}{%
  \bibinfo {author} {\bibfnamefont{W.}~\bibnamefont{Feng}}, \bibinfo {author}
  {\bibfnamefont{D.}~\bibnamefont{Xiao}}, \bibinfo {author}
  {\bibfnamefont{J.}~\bibnamefont{Ding}},\ and\ \bibinfo {author}
  {\bibfnamefont{Y.}~\bibnamefont{Yao}},\ }%
  \bibfield{journal}{%
  \bibinfo {journal} {Phys. Rev. Lett.}\ }%
  \textbf{\bibinfo {volume} {106}},\ \bibinfo {pages} {016402} (\bibinfo {year}
  {2011})%
  \bibAnnoteFile{NoStop}{FengPRL2010gu}%
\bibitem{WadaPRB2011fu}%
  \BibitemOpen
  \bibfield{author}{%
  \bibinfo {author} {\bibfnamefont{M.}~\bibnamefont{Wada}}, \bibinfo {author}
  {\bibfnamefont{S.}~\bibnamefont{Murakami}}, \bibinfo {author}
  {\bibfnamefont{F.}~\bibnamefont{Freimuth}},\ and\ \bibinfo {author}
  {\bibfnamefont{G.}~\bibnamefont{Bihlmayer}},\ }%
  \bibfield{journal}{%
  \bibinfo {journal} {Phys. Rev. B}\ }%
  \textbf{\bibinfo {volume} {83}},\ \bibinfo {pages} {121310} (\bibinfo {year}
  {2011})%
  \bibAnnoteFile{NoStop}{WadaPRB2011fu}%
\bibitem{DaiPRB2008ji}%
  \BibitemOpen
  \bibfield{author}{%
  \bibinfo {author} {\bibfnamefont{X.}~\bibnamefont{Dai}}, \bibinfo {author}
  {\bibfnamefont{T.~L.}\ \bibnamefont{Hughes}}, \bibinfo {author}
  {\bibfnamefont{X.-L.}\ \bibnamefont{Qi}}, \bibinfo {author}
  {\bibfnamefont{Z.}~\bibnamefont{Fang}},\ and\ \bibinfo {author}
  {\bibfnamefont{S.-C.}\ \bibnamefont{Zhang}},\ }%
  \bibfield{journal}{%
  \bibinfo {journal} {Phys. Rev. B}\ }%
  \textbf{\bibinfo {volume} {77}},\ \bibinfo {pages} {125319} (\bibinfo {year}
  {2008})%
  \bibAnnoteFile{NoStop}{DaiPRB2008ji}%
\bibitem{Virot2010by}%
  \BibitemOpen
  \bibfield{author}{%
  \bibinfo {author} {\bibfnamefont{F.}~\bibnamefont{Virot}}, \bibinfo {author}
  {\bibfnamefont{R.}~\bibnamefont{Hayn}}, \bibinfo {author}
  {\bibfnamefont{M.}~\bibnamefont{Richter}},\ and\ \bibinfo {author}
  {\bibfnamefont{J.}~\bibnamefont{van~den Brink}},\ \bibinfo {pages}
  {arXiv:1105.0501v1}}%
   (\bibinfo {year} {2011})%
  \bibAnnoteFile{NoStop}{Virot2010by}%
\bibitem{Soluyanov2011gy}%
  \BibitemOpen
  \bibfield{author}{%
  \bibinfo {author} {\bibfnamefont{A.~A.}\ \bibnamefont{Soluyanov}}\ and\
  \bibinfo {author} {\bibfnamefont{D.}~\bibnamefont{Vanderbilt}},\ \bibinfo
  {pages} {arXiv:1102.5600v1}}%
   (\bibinfo {year} {2011})%
  \bibAnnoteFile{NoStop}{Soluyanov2011gy}%
\bibitem{Yu2011re}%
  \BibitemOpen
  \bibfield{author}{%
  \bibinfo {author} {\bibfnamefont{R.}~\bibnamefont{Yu}}, \bibinfo {author}
  {\bibfnamefont{X.-L.}\ \bibnamefont{Qi}}, \bibinfo {author}
  {\bibfnamefont{A.}~\bibnamefont{Bernevig}}, \bibinfo {author}
  {\bibfnamefont{Z.}~\bibnamefont{Fang}},\ and\ \bibinfo {author}
  {\bibfnamefont{X.}~\bibnamefont{Dai}},\ \bibinfo {pages}
  {arXiv:1101.2011v1}}%
   (\bibinfo {year} {2011})%
  \bibAnnoteFile{NoStop}{Yu2011re}%
\bibitem{Ringel2010vo}%
  \BibitemOpen
  \bibfield{author}{%
  \bibinfo {author} {\bibfnamefont{Z.}~\bibnamefont{Ringel}}\ and\ \bibinfo
  {author} {\bibfnamefont{E.}~\bibnamefont{Kraus}},\ \bibinfo {pages}
  {arXiv:1010.5357v2}}%
   (\bibinfo {year} {2010})%
  \bibAnnoteFile{NoStop}{Ringel2010vo}%
\bibitem{QiPRB2008ng}%
  \BibitemOpen
  \bibfield{author}{%
  \bibinfo {author} {\bibfnamefont{X.-L.}\ \bibnamefont{Qi}}, \bibinfo {author}
  {\bibfnamefont{T.~L.}\ \bibnamefont{Hughes}},\ and\ \bibinfo {author}
  {\bibfnamefont{S.-C.}\ \bibnamefont{Zhang}},\ }%
  \bibfield{journal}{%
  \bibinfo {journal} {Phys. Rev. B}\ }%
  \textbf{\bibinfo {volume} {78}},\ \bibinfo {pages} {195424} (\bibinfo {year}
  {2008})%
  \bibAnnoteFile{NoStop}{QiPRB2008ng}%
\bibitem{Wang2010xs}%
  \BibitemOpen
  \bibfield{author}{%
  \bibinfo {author} {\bibfnamefont{Z.}~\bibnamefont{Wang}}, \bibinfo {author}
  {\bibfnamefont{X.}~\bibnamefont{Qi}},\ and\ \bibinfo {author}
  {\bibfnamefont{S.-C.}\ \bibnamefont{Zhang}},\ }%
  \bibfield{journal}{%
  \bibinfo {journal} {New J. Phys.}\ }%
  \textbf{\bibinfo {volume} {12}},\ \bibinfo {pages} {065007} (\bibinfo {year}
  {2010})%
  \bibAnnoteFile{NoStop}{Wang2010xs}%
\bibitem{CohPRB2011rt}%
  \BibitemOpen
  \bibfield{author}{%
  \bibinfo {author} {\bibfnamefont{S.}~\bibnamefont{Coh}}, \bibinfo {author}
  {\bibfnamefont{D.}~\bibnamefont{Vanderbilt}}, \bibinfo {author}
  {\bibfnamefont{A.}~\bibnamefont{Malashevich}},\ and\ \bibinfo {author}
  {\bibfnamefont{I.}~\bibnamefont{Souza}},\ }%
  \bibfield{journal}{%
  \bibinfo {journal} {Phys. Rev. B}\ }%
  \textbf{\bibinfo {volume} {83}},\ \bibinfo {pages} {085108} (\bibinfo {year}
  {2011})%
  \bibAnnoteFile{NoStop}{CohPRB2011rt}%
\bibitem{Qi:2008cg}%
  \BibitemOpen
  \bibfield{author}{%
  \bibinfo {author} {\bibfnamefont{X.-L.}\ \bibnamefont{Qi}}, \bibinfo {author}
  {\bibfnamefont{T.~L.}\ \bibnamefont{Hughes}},\ and\ \bibinfo {author}
  {\bibfnamefont{S.-C.}\ \bibnamefont{Zhang}},\ }%
  \bibfield{journal}{%
  \bibinfo {journal} {Phys. Rev. B}\ }%
  \textbf{\bibinfo {volume} {78}},\ \bibinfo {pages} {195424} (\bibinfo {year}
  {2008})%
  \bibAnnoteFile{NoStop}{Qi:2008cg}%
\bibitem{Hughes2010gh}%
  \BibitemOpen
  \bibfield{author}{%
  \bibinfo {author} {\bibfnamefont{T.}~\bibnamefont{Hughes}}, \bibinfo {author}
  {\bibfnamefont{E.}~\bibnamefont{Prodan}},\ and\ \bibinfo {author}
  {\bibfnamefont{B.~A.}\ \bibnamefont{Bernevig}},\ \bibinfo {pages}
  {arxiv:1010.4508}}%
   (\bibinfo {year} {2010})%
  \bibAnnoteFile{NoStop}{Hughes2010gh}%
\bibitem{simon1983}%
  \BibitemOpen
  \bibfield{author}{%
  \bibinfo {author} {\bibfnamefont{B.}~\bibnamefont{Simon}},\ }%
  \bibfield{journal}{%
  \bibinfo {journal} {Phys. Rev. Lett.}\ }%
  \textbf{\bibinfo {volume} {51}},\ \bibinfo {pages} {2167} (\bibinfo {year}
  {1983})%
  \bibAnnoteFile{NoStop}{simon1983}%
\bibitem{Nenciu:1981kx}%
  \BibitemOpen
  \bibfield{author}{%
  \bibinfo {author} {\bibfnamefont{G.}~\bibnamefont{Nenciu}},\ }%
  \bibfield{journal}{%
  \bibinfo {journal} {Comm. Math. Phys.}\ }%
  \textbf{\bibinfo {volume} {82}},\ \bibinfo {pages} {121} (\bibinfo {year}
  {1981})%
  \bibAnnoteFile{NoStop}{Nenciu:1981kx}%
\bibitem{wilczek:1984bs}%
  \BibitemOpen
  \bibfield{author}{%
  \bibinfo {author} {\bibfnamefont{F.}~\bibnamefont{Wilczek}}\ and\ \bibinfo
  {author} {\bibfnamefont{A.}~\bibnamefont{Zee}},\ }%
  \bibfield{journal}{%
  \bibinfo {journal} {Phys. Rev. Lett.}\ }%
  \textbf{\bibinfo {volume} {52}},\ \bibinfo {pages} {2111} (\bibinfo {year}
  {1984})%
  \bibAnnoteFile{NoStop}{wilczek:1984bs}%
\bibitem{Prodan:2009hg}%
  \BibitemOpen
  \bibfield{author}{%
  \bibinfo {author} {\bibfnamefont{E.}~\bibnamefont{Prodan}}\ and\ \bibinfo
  {author} {\bibfnamefont{F.~D.~M.}\ \bibnamefont{Haldane}},\ }%
  \bibfield{journal}{%
  \bibinfo {journal} {Phys. Rev. B}\ }%
  \textbf{\bibinfo {volume} {80}},\ \bibinfo {pages} {115121} (\bibinfo {year}
  {2009})%
  \bibAnnoteFile{NoStop}{Prodan:2009hg}%
\bibitem{SimonTr2005xu}%
  \BibitemOpen
  \bibfield{author}{%
  \bibinfo {author} {\bibfnamefont{B.}~\bibnamefont{Simon}},\ }%
  \emph{\bibinfo {title} {Trace ideals and their applications}},\ \bibinfo
  {series} {Mathematical Surveys and Monographs}, Vol.\ \bibinfo {volume}
  {120}\ (\bibinfo {publisher} {Americal Mathematical Society, Providence},\
  \bibinfo {year} {2005})%
  \bibAnnoteFile{NoStop}{SimonTr2005xu}%
\bibitem{Roy2010nj}%
  \BibitemOpen
  \bibfield{author}{%
  \bibinfo {author} {\bibfnamefont{R.}~\bibnamefont{Roy}},\ }%
  \bibfield{journal}{%
  \bibinfo {journal} {New J. Phys.}\ }%
  \textbf{\bibinfo {volume} {12}},\ \bibinfo {pages} {065009} (\bibinfo {year}
  {2010})%
  \bibAnnoteFile{NoStop}{Roy2010nj}%
\bibitem{Yamakage2010xr}%
  \BibitemOpen
  \bibfield{author}{%
  \bibinfo {author} {\bibfnamefont{A.}~\bibnamefont{Yamakage}}, \bibinfo
  {author} {\bibfnamefont{K.}~\bibnamefont{Nomura}}, \bibinfo {author}
  {\bibfnamefont{K.~I.}\ \bibnamefont{Imura}},\ and\ \bibinfo {author}
  {\bibfnamefont{Y.}~\bibnamefont{Kuramoto}},\ }%
  \bibfield{journal}{%
  \bibinfo {journal} {J. Phys. Soc. Jpn.}\ }%
  \textbf{\bibinfo {volume} {80}},\ \bibinfo {pages} {053703} (\bibinfo {year}
  {2011})%
  \bibAnnoteFile{NoStop}{Yamakage2010xr}%
\bibitem{Prodan2011vy}%
  \BibitemOpen
  \bibfield{author}{%
  \bibinfo {author} {\bibfnamefont{E.}~\bibnamefont{Prodan}},\ }%
  \bibfield{journal}{%
  \bibinfo {journal} {Phys. Rev. B}\ }%
  \textbf{\bibinfo {volume} {83}},\ \bibinfo {pages} {195119} (\bibinfo {year}
  {2011})%
  \bibAnnoteFile{NoStop}{Prodan2011vy}%
\bibitem{LiuPRB2010xf}%
  \BibitemOpen
  \bibfield{author}{%
  \bibinfo {author} {\bibfnamefont{C.-X.}\ \bibnamefont{Liu}}, \bibinfo
  {author} {\bibfnamefont{X.-L.}\ \bibnamefont{Qi}}, \bibinfo {author}
  {\bibfnamefont{H.}~\bibnamefont{Zhang}}, \bibinfo {author}
  {\bibfnamefont{X.}~\bibnamefont{Dai}}, \bibinfo {author}
  {\bibfnamefont{Z.}~\bibnamefont{Fang}},\ and\ \bibinfo {author}
  {\bibfnamefont{S.-C.}\ \bibnamefont{Zhang}},\ }%
  \bibfield{journal}{%
  \bibinfo {journal} {Phys. Rev. B}\ }%
  \textbf{\bibinfo {volume} {82}},\ \bibinfo {pages} {045122} (\bibinfo {year}
  {2010})%
  \bibAnnoteFile{NoStop}{LiuPRB2010xf}%
\bibitem{Qi2010hg}%
  \BibitemOpen
  \bibfield{author}{%
  \bibinfo {author} {\bibfnamefont{X.~L.}\ \bibnamefont{Qi}}\ and\ \bibinfo
  {author} {\bibfnamefont{S.-C.}\ \bibnamefont{Zhang}},\ }%
  \bibfield{journal}{%
  \bibinfo {journal} {arXiv:1008.2026v1}}%
   (\bibinfo {year} {2010})%
  \bibAnnoteFile{NoStop}{Qi2010hg}%
\bibitem{ProdanJPhysA2010xk}%
  \BibitemOpen
  \bibfield{author}{%
  \bibinfo {author} {\bibfnamefont{E.}~\bibnamefont{Prodan}},\ }%
  \bibfield{journal}{%
  \bibinfo {journal} {J. Phys. A: Math. Theor.}\ }%
  \textbf{\bibinfo {volume} {44}},\ \bibinfo {pages} {113001} (\bibinfo {year}
  {2011})%
  \bibAnnoteFile{NoStop}{ProdanJPhysA2010xk}%
\bibitem{BergPRE2011vy}%
  \BibitemOpen
  \bibfield{author}{%
  \bibinfo {author} {\bibfnamefont{N.}~\bibnamefont{Berg}}, \bibinfo {author}
  {\bibfnamefont{K.}~\bibnamefont{Joel}}, \bibinfo {author}
  {\bibfnamefont{M.}~\bibnamefont{Koolyk}},\ and\ \bibinfo {author}
  {\bibfnamefont{E.}~\bibnamefont{Prodan}},\ }%
  \bibfield{journal}{%
  \bibinfo {journal} {Phys. Rev. E}\ }%
  \textbf{\bibinfo {volume} {83}},\ \bibinfo {pages} {021913} (\bibinfo {year}
  {2011})%
  \bibAnnoteFile{NoStop}{BergPRE2011vy}%
\end{thebibliography}

%

\end{document}